\documentclass[conference]{IEEEtran}
\IEEEoverridecommandlockouts

\UseRawInputEncoding

\usepackage{cite}
\usepackage{amsmath,amssymb,amsfonts}
\usepackage{algorithmic}
\usepackage{graphicx}
\usepackage{textcomp}
\usepackage{xcolor}
\usepackage{amsthm}
\usepackage{amsmath}
\usepackage{amssymb}
\usepackage{booktabs}
\usepackage{amsfonts}
\usepackage{lipsum}
\usepackage{mathtools}
\usepackage{threeparttable}
\usepackage{tablefootnote}
\usepackage[hyphens]{url}
\usepackage[hidelinks]{hyperref}
\usepackage{algorithmic, algorithm}
\usepackage{tabularx}
\usepackage{subfig}
\usepackage{mwe}
\usepackage{ragged2e}
\usepackage{dblfloatfix}
\usepackage{footmisc}

\makeatletter
\newcommand\fs@betterruled{%
  \def\@fs@cfont{\bfseries}\let\@fs@capt\floatc@ruled
  \def\@fs@pre{\vspace*{5pt}\hrule height.8pt depth0pt \kern2pt}%
  \def\@fs@post{\kern2pt\hrule\relax}%
  \def\@fs@mid{\kern2pt\hrule\kern2pt}%
  \let\@fs@iftopcapt\iftrue}
\floatstyle{betterruled}
\restylefloat{algorithm}
\makeatother

\definecolor{custom_blue}{HTML}{1F77B4}
\definecolor{custom_orange}{HTML}{FF7F0E}
\definecolor{custom_green}{HTML}{2CA02C}

\usepackage{tikz}
\usepackage{pgfplots}
\pgfplotsset{compat=1.17}

\newcommand{\blue}{\raisebox{2pt}{\tikz{\draw[custom_blue, solid, line width=2.3pt](0,0) -- (5mm,0);}}}
\newcommand{\orange}{\raisebox{2pt}{\tikz{\draw[custom_orange, solid, line width=2.3pt](0,0) -- (5mm,0);}}}
\newcommand{\green}{\raisebox{2pt}{\tikz{\draw[custom_green, solid, line width=2.3pt](0,0) -- (5mm,0);}}}

\usepackage[nolist,printonlyused]{acronym} 
\def\BibTeX{{\rm B\kern-.05em{\sc i\kern-.025em b}\kern-.08em
    T\kern-.1667em\lower.7ex\hbox{E}\kern-.125emX}}
\begin{document}

\begin{acronym}
  \acro{2G}{Second Generation}
  \acro{3G}{3$^\text{rd}$~Generation}
  \acro{3GPP}{3$^\text{rd}$~Generation Partnership Project}
  \acro{4G}{4$^\text{th}$~Generation}
  \acro{5G}{5$^\text{th}$~Generation}
  \acro{AA}{Antenna Array}
  \acro{AC}{Admission Control}
  \acro{AD}{Attack-Decay}
  \acro{ADSL}{Asymmetric Digital Subscriber Line}
	\acro{AHW}{Alternate Hop-and-Wait}
  \acro{AMC}{Adaptive Modulation and Coding}
	\acro{AP}{access point}
  \acro{AoI}{age of information}
  \acro{APA}{Adaptive Power Allocation}
  \acro{AR}{autoregressive}
  \acro{ARMA}{Autoregressive Moving Average}
  \acro{ATES}{Adaptive Throughput-based Efficiency-Satisfaction Trade-Off}
  \acro{AWGN}{additive white Gaussian noise}
  \acro{BB}{Branch and Bound}
  \acro{BD}{Block Diagonalization}
  \acro{BER}{bit error rate}
  \acro{BF}{Best Fit}
  \acro{BLER}{BLock Error Rate}
  \acro{BPC}{Binary power control}
  \acro{BPSK}{binary phase-shift keying}
  \acro{BPA}{Best \ac{PDPR} Algorithm}
  \acro{BRA}{Balanced Random Allocation}
  \acro{BS}{base station}
  \acro{CAP}{Combinatorial Allocation Problem}
  \acro{CAPEX}{Capital Expenditure}
  \acro{CB}{codebook}
  \acro{CBF}{Coordinated Beamforming}
  \acro{CBR}{Constant Bit Rate}
  \acro{CBS}{Class Based Scheduling}
  \acro{CC}{Congestion Control}
  \acro{CDF}{Cumulative Distribution Function}
  \acro{CDMA}{Code-Division Multiple Access}
  \acro{CL}{Closed Loop}
  \acro{CLPC}{Closed Loop Power Control}
  \acro{CNR}{Channel-to-Noise Ratio}
  \acro{CPA}{Cellular Protection Algorithm}
  \acro{CPICH}{Common Pilot Channel}
  \acro{CoMP}{Coordinated Multi-Point}
  \acro{CQI}{Channel Quality Indicator}
  \acro{CRLB}{Cramér-Rao Lower Bound}
  \acro{CRM}{Constrained Rate Maximization}
	\acro{CRN}{Cognitive Radio Network}
  \acro{CS}{Coordinated Scheduling}
  \acro{CSI}{channel state information}
  \acro{CSIR}{channel state information at the receiver}
  \acro{CSIT}{channel state information at the transmitter}
  \acro{CUE}{cellular user equipment}
  \acro{D2D}{device-to-device}
  \acro{DCA}{Dynamic Channel Allocation}
  \acro{DE}{Differential Evolution}
  \acro{DFT}{Discrete Fourier Transform}
  \acro{DISCOVER}{Deep Directed Intrinsically Motivated Exploration}
  \acro{DIST}{Distance}
  \acro{DL}{downlink}
  \acro{DMA}{Double Moving Average}
	\acro{DMRS}{Demodulation Reference Signal}
  \acro{D2DM}{D2D Mode}
  \acro{DMS}{D2D Mode Selection}
  \acro{DPC}{Dirty Paper Coding}
  \acro{DRA}{Dynamic Resource Assignment}
  \acro{DRL}{deep reinforcement learning}
  \acro{DDPG}{Deep Deterministic Policy Gradient}
  \acro{RL}{reinforcement learning}
  \acro{SAC}{Soft Actor-Critic}
  \acro{DSA}{Dynamic Spectrum Access}
  \acro{DSM}{Delay-based Satisfaction Maximization}
  \acro{DQN}{Deeq Q-Network}
  \acro{ECC}{Electronic Communications Committee}
  \acro{ECRB}{expectation of the conditional Cramér-Rao bound}
  \acro{EFLC}{Error Feedback Based Load Control}
  \acro{EI}{Efficiency Indicator}
  \acro{eNB}{Evolved Node B}
  \acro{EPA}{Equal Power Allocation}
  \acro{EPC}{Evolved Packet Core}
  \acro{EPS}{Evolved Packet System}
  \acro{E-UTRAN}{Evolved Universal Terrestrial Radio Access Network}
  \acro{ES}{Exhaustive Search}
  \acro{FDD}{frequency division duplexing}
  \acro{FDM}{Frequency Division Multiplexing}
  \acro{FER}{Frame Erasure Rate}
  \acro{FF}{Fast Fading}
  \acro{FIM}{Fisher Information Matrix}
  \acro{FSB}{Fixed Switched Beamforming}
  \acro{FST}{Fixed SNR Target}
  \acro{FTP}{File Transfer Protocol}
  \acro{GA}{Genetic Algorithm}
  \acro{GBR}{Guaranteed Bit Rate}
  \acro{GLR}{Gain to Leakage Ratio}
  \acro{GOS}{Generated Orthogonal Sequence}
  \acro{GPL}{GNU General Public License}
  \acro{GRP}{Grouping}
  \acro{HARQ}{Hybrid Automatic Repeat Request}
  \acro{HMS}{Harmonic Mode Selection}
  \acro{HOL}{Head Of Line}
  \acro{HSDPA}{High-Speed Downlink Packet Access}
  \acro{HSPA}{High Speed Packet Access}
  \acro{HTTP}{HyperText Transfer Protocol}
  \acro{ICMP}{Internet Control Message Protocol}
  \acro{ICI}{Intercell Interference}
  \acro{ID}{Identification}
  \acro{IETF}{Internet Engineering Task Force}
  \acro{ILP}{Integer Linear Program}
  \acro{JRAPAP}{Joint RB Assignment and Power Allocation Problem}
  \acro{UID}{Unique Identification}
  \acro{i.i.d.}{independent and identically distributed}
  \acro{IIR}{Infinite Impulse Response}
  \acro{ILP}{Integer Linear Problem}
  \acro{IMT}{International Mobile Telecommunications}
  \acro{INV}{Inverted Norm-based Grouping}
	\acro{IoT}{Internet of Things}
  \acro{IP}{Internet Protocol}
  \acro{IPv6}{Internet Protocol Version 6}
  \acro{IRS}{intelligent reflecting surface}
  \acro{ISD}{Inter-Site Distance}
  \acro{ISI}{Inter Symbol Interference}
  \acro{ITU}{International Telecommunication Union}
  \acro{JOAS}{Joint Opportunistic Assignment and Scheduling}
  \acro{JOS}{Joint Opportunistic Scheduling}
  \acro{JP}{Joint Processing}
	\acro{JS}{Jump-Stay}
    \acro{KF}{Kalman filter}
  \acro{KKT}{Karush-Kuhn-Tucker}
  \acro{L3}{Layer-3}
  \acro{LAC}{Link Admission Control}
  \acro{LA}{Link Adaptation}
  \acro{LC}{Load Control}
  \acro{LMMSE}{linear minimum mean squared error}
  \acro{LOS}{Line of Sight}
  \acro{LP}{Linear Programming}
  \acro{LS}{least squares}
  \acro{LTE}{Long Term Evolution}
  \acro{LTE-A}{LTE-Advanced}
  \acro{LTE-Advanced}{Long Term Evolution Advanced}
  \acro{M2M}{Machine-to-Machine}
  \acro{MAC}{Medium Access Control}
  \acro{MANET}{Mobile Ad hoc Network}
  \acro{MC}{Modular Clock}
  \acro{MCS}{Modulation and Coding Scheme}
  \acro{MDB}{Measured Delay Based}
  \acro{MDI}{Minimum D2D Interference}
  \acro{MF}{Matched Filter}
  \acro{MG}{Maximum Gain}
  \acro{MH}{Multi-Hop}
  \acro{MIMO}{multiple input multiple output}
  \acro{MINLP}{Mixed Integer Nonlinear Programming}
  \acro{MIP}{Mixed Integer Programming}
  \acro{MISO}{Multiple Input Single Output}
  \acro{ML}{maximum likelihood}
  \acro{MLWDF}{Modified Largest Weighted Delay First}
  \acro{MME}{Mobility Management Entity}
  \acro{MMSE}{minimum mean squared error}
  \acro{MOS}{Mean Opinion Score}
  \acro{MPF}{Multicarrier Proportional Fair}
  \acro{MRA}{Maximum Rate Allocation}
  \acro{MR}{Maximum Rate}
  \acro{MRC}{maximum ratio combining}
  \acro{MRT}{Maximum Ratio Transmission}
  \acro{MRUS}{Maximum Rate with User Satisfaction}
  \acro{MS}{mobile station}
  \acro{MSE}{mean squared error}
  \acro{MSI}{Multi-Stream Interference}
  \acro{MTC}{Machine-Type Communication}
  \acro{MTSI}{Multimedia Telephony Services over IMS}
  \acro{MTSM}{Modified Throughput-based Satisfaction Maximization}
  \acro{MU-MIMO}{multiuser multiple input multiple output}
  \acro{MU-MISO}{multiuser multiple input single output}
  \acro{MU}{multi-user}
  \acro{NAS}{Non-Access Stratum}
  \acro{NB}{Node B}
  \acro{NE}{Nash equilibrium}
  \acro{NCL}{Neighbor Cell List}
  \acro{NLP}{Nonlinear Programming}
  \acro{NLOS}{Non-Line of Sight}
  \acro{NMSE}{normalized mean squared error}
  \acro{NOMA}{non-orthogonal multiple access}
  \acro{NORM}{Normalized Projection-based Grouping}
  \acro{NP}{Non-Polynomial Time}
  \acro{NR}{New Radio}
  \acro{NRT}{Non-Real Time}
  \acro{NSPS}{National Security and Public Safety Services}
  \acro{O2I}{Outdoor to Indoor}
  \acro{OFDMA}{orthogonal frequency division multiple access}
  \acro{OFDM}{orthogonal frequency division multiplexing}
  \acro{OFPC}{Open Loop with Fractional Path Loss Compensation}
	\acro{O2I}{Outdoor-to-Indoor}
  \acro{OL}{Open Loop}
  \acro{OLPC}{Open-Loop Power Control}
  \acro{OL-PC}{Open-Loop Power Control}
  \acro{OPEX}{Operational Expenditure}
  \acro{ORB}{Orthogonal Random Beamforming}
  \acro{JO-PF}{Joint Opportunistic Proportional Fair}
  \acro{OSI}{Open Systems Interconnection}
  \acro{PAIR}{D2D Pair Gain-based Grouping}
  \acro{PAPR}{Peak-to-Average Power Ratio}
  \acro{P2P}{Peer-to-Peer}
  \acro{PC}{Power Control}
  \acro{PCI}{Physical Cell ID}
  \acro{PDF}{Probability Density Function}
  \acro{PDPR}{pilot-to-data power ratio}
  \acro{PER}{Packet Error Rate}
  \acro{PF}{Proportional Fair}
  \acro{P-GW}{Packet Data Network Gateway}
  \acro{PL}{Pathloss}
  \acro{PPR}{pilot power ratio}
  \acro{PRB}{physical resource block}
  \acro{PROJ}{Projection-based Grouping}
  \acro{ProSe}{Proximity Services}
  \acro{PS}{Packet Scheduling}
  \acro{PSAM}{pilot symbol assisted modulation}
  \acro{PSO}{Particle Swarm Optimization}
  \acro{PZF}{Projected Zero-Forcing}
  \acro{QAM}{Quadrature Amplitude Modulation}
  \acro{QoS}{Quality of Service}
  \acro{QPSK}{Quadri-Phase Shift Keying}
  \acro{RAISES}{Reallocation-based Assignment for Improved Spectral Efficiency and Satisfaction}
  \acro{RAN}{Radio Access Network}
  \acro{RA}{Resource Allocation}
  \acro{RAT}{Radio Access Technology}
  \acro{RATE}{Rate-based}
  \acro{RB}{resource block}
  \acro{RBG}{Resource Block Group}
  \acro{REF}{Reference Grouping}
  \acro{RIS}{reconfigurable intelligent surface}
  \acro{RLC}{Radio Link Control}
  \acro{RM}{Rate Maximization}
  \acro{RNC}{Radio Network Controller}
  \acro{RND}{Random Grouping}
  \acro{RRA}{Radio Resource Allocation}
  \acro{RRM}{Radio Resource Management}
  \acro{RSCP}{Received Signal Code Power}
  \acro{RSRP}{Reference Signal Receive Power}
  \acro{RSRQ}{Reference Signal Receive Quality}
  \acro{RR}{Round Robin}
  \acro{RRC}{Radio Resource Control}
  \acro{RSSI}{Received Signal Strength Indicator}
  \acro{RT}{Real Time}
  \acro{RU}{Resource Unit}
  \acro{RUNE}{RUdimentary Network Emulator}
  \acro{RV}{Random Variable}
  \acro{SAC}{Soft Actor-Critic}
  \acro{SCM}{Spatial Channel Model}
  \acro{SC-FDMA}{Single Carrier - Frequency Division Multiple Access}
  \acro{SD}{Soft Dropping}
  \acro{S-D}{Source-Destination}
  \acro{SDPC}{Soft Dropping Power Control}
  \acro{SDMA}{Space-Division Multiple Access}
  \acro{SE}{spectral efficiency}
  \acro{SER}{Symbol Error Rate}
  \acro{SES}{Simple Exponential Smoothing}
  \acro{S-GW}{Serving Gateway}
  \acro{SINR}{signal-to-interference-plus-noise ratio}
  \acro{SI}{Satisfaction Indicator}
  \acro{SIP}{Session Initiation Protocol}
  \acro{SISO}{single input single output}
  \acro{SIMO}{single input multiple output}
  \acro{SIR}{signal-to-interference ratio}
  \acro{SLNR}{Signal-to-Leakage-plus-Noise Ratio}
  \acro{SMA}{Simple Moving Average}
  \acro{SNR}{signal-to-noise ratio}
  \acro{SORA}{Satisfaction Oriented Resource Allocation}
  \acro{SORA-NRT}{Satisfaction-Oriented Resource Allocation for Non-Real Time Services}
  \acro{SORA-RT}{Satisfaction-Oriented Resource Allocation for Real Time Services}
  \acro{SPF}{Single-Carrier Proportional Fair}
  \acro{SRA}{Sequential Removal Algorithm}
  \acro{SRS}{Sounding Reference Signal}
  \acro{SU-MIMO}{single-user multiple input multiple output}
  \acro{SU}{Single-User}
  \acro{SVD}{Singular Value Decomposition}
  \acro{TCP}{Transmission Control Protocol}
  \acro{TDD}{time division duplexing}
  \acro{TDMA}{Time Division Multiple Access}
  \acro{TETRA}{Terrestrial Trunked Radio}
  \acro{TP}{Transmit Power}
  \acro{TPC}{Transmit Power Control}
  \acro{TTI}{Transmission Time Interval}
  \acro{TTR}{Time-To-Rendezvous}
  \acro{TSM}{Throughput-based Satisfaction Maximization}
  \acro{TU}{Typical Urban}
  \acro{UAV}{unmanned aerial vehicle}
  \acro{UE}{user equipment}
  \acro{UEPS}{Urgency and Efficiency-based Packet Scheduling}
  \acro{UL}{uplink}
  \acro{UMTS}{Universal Mobile Telecommunications System}
  \acro{URI}{Uniform Resource Identifier}
  \acro{URM}{Unconstrained Rate Maximization}
  \acro{UT}{user terminal}
  \acro{VAR}{vector autoregressive}
  \acro{VR}{Virtual Resource}
  \acro{VoIP}{Voice over IP}
  \acro{WAN}{Wireless Access Network}
  \acro{WCDMA}{Wideband Code Division Multiple Access}
  \acro{WF}{Water-filling}
  \acro{WiMAX}{Worldwide Interoperability for Microwave Access}
  \acro{WINNER}{Wireless World Initiative New Radio}
  \acro{WLAN}{Wireless Local Area Network}
  \acro{WMPF}{Weighted Multicarrier Proportional Fair}
  \acro{WPF}{Weighted Proportional Fair}
  \acro{WSN}{Wireless Sensor Network}
  \acro{WSS}{wide-sense stationary}
  \acro{WWW}{World Wide Web}
  \acro{XIXO}{(Single or Multiple) Input (Single or Multiple) Output}
  \acro{ZF}{zero-forcing}
  \acro{ZMCSCG}{Zero Mean Circularly Symmetric Complex Gaussian}
\end{acronym}

\title{Deep Reinforcement Learning Based Joint Downlink Beamforming and RIS Configuration in RIS-aided MU-MISO Systems Under Hardware Impairments and Imperfect CSI}
\author{\IEEEauthorblockN{Baturay Saglam\IEEEauthorrefmark{1}, \emph{Student Member, IEEE}, Doga Gurgunoglu\IEEEauthorrefmark{2}, \emph{Student Member, IEEE,}} Suleyman S. Kozat\IEEEauthorrefmark{1}, \emph{Senior Member, IEEE}
\IEEEauthorblockA{\IEEEauthorrefmark{1}Department of Electrical and Electronics Engineering, Bilkent University, Ankara 06800, Turkey\\\IEEEauthorrefmark{2}Division of Decision and Control Systems, KTH Royal Institute of Technology, Stockholm 100 44, Sweden}
\thanks{This study is supported by Turk Telekom within the framework of the 5G and Beyond Joint Graduate Support Programme coordinated by Information and Communication Technologies Authority and the EU Horizon 2020 MSCA-ITN-METAWIRELESS, Grant Agreement 956256.}
\vspace{-0.7cm}}


\maketitle

\begin{abstract}
We introduce a novel \ac{DRL} approach to jointly optimize transmit beamforming and \ac{RIS} phase shifts in a \ac{MU-MISO} system to maximize the sum downlink rate under the phase-dependent reflection amplitude model. Our approach addresses the challenge of imperfect \ac{CSI} and hardware impairments by considering a practical \ac{RIS} amplitude model. We compare the performance of our approach against a vanilla \ac{DRL} agent in two scenarios: perfect \ac{CSI} and phase-dependent \ac{RIS} amplitudes, and mismatched \ac{CSI} and ideal \ac{RIS} reflections. The results demonstrate that the proposed framework significantly outperforms the vanilla \ac{DRL} agent under mismatch and approaches the golden standard. Our contributions include modifications to the \ac{DRL} approach to address the joint design of transmit beamforming and phase shifts and the phase-dependent amplitude model. To the best of our knowledge, our method is the first \ac{DRL}-based approach for the phase-dependent reflection amplitude model in \ac{RIS}-aided \ac{MU-MISO} systems. Our findings in this study highlight the potential of our approach as a promising solution to overcome hardware impairments in \ac{RIS}-aided wireless communication systems.
\end{abstract}


\section{Introduction}
\ac{RIS} is among the emerging technologies explored for next-generation wireless communication systems \cite{pdaModel}. An \ac{RIS} consists of multiple reflecting elements with sub-wavelength spacing whose impedances are adjusted to induce desired phase shifts on incident waves before they are reflected. This enables the manipulation of multipath interference at the receiver \cite{sinan_hoca}. However, depending on the \ac{RIS} hardware, the incident wave is attenuated depending on the applied phase shifts to the individual elements, resulting in \emph{phase-dependent reflection amplitudes} \cite{pdaModel}. Such phenomenon causes significant performance losses \cite{sinan_hoca}.

The non-linear model in \cite{pdaModel} renders the already complex optimization-based approaches impractical \cite{pose}. An alternative to such methods, deep reinforcement learning (DRL), has become a widely-studied machine learning (ML) approach for \ac{RIS}-aided wireless systems such as \ac{NOMA} downlink systems \cite{97}, millimeter wave communications \cite{96}, vehicular communications and trajectory optimization \cite{101,100,103}, and the transmit beamforming and phase shifts design \cite{91,93,94,95,98,99,101,jointBFRIS}. In \cite{97}, \ac{RIS} phase shifts are adjusted using the \ac{DDPG} algorithm \cite{ddpg}. In \cite{96}, the joint design of the downlink beamforming matrix and the \ac{RIS} phase shifts are considered under ideal \ac{RIS} reflections. In \cite{jointBFRIS}, the aforementioned optimization is performed under individual users' \ac{SINR} constraints, as opposed to maximizing the downlink sum rate, which is susceptible to maximizing the sum rate by significantly lowering certain users' individual rates. In \cite{wcl2023ofdm}, a \ac{DQN}-based framework is proposed to maximize the \ac{SE} of the downlink of an \ac{OFDM} communication system with a low-resolution \ac{RIS}. While the prior applications of \ac{DRL} to \ac{RIS}-aided systems assumed ideal reflections and perfect \ac{CSI}, a \ac{DRL} application considering \ac{RIS} hardware impairments does not exist to the best of our knowledge.

In this paper, we study the joint design of transmit beamforming and phase shifts for \ac{RIS}-aided multi-user multiple input single output (MU-MISO) systems through a \ac{DRL} approach. Our objective is to maximize the sum downlink rate of the users under the phase-dependent amplitude model. Since phase-dependent amplitudes make the system more complex, we only consider \ac{DRL}-based approaches in our study. The main contributions of this study can be summarized as follows: 

\begin{itemize}
\item We present novel modifications for the application of \ac{DRL} to RIS-aided systems, which address two critical aspects of \emph{non-episodic} tasks, e.g., the joint design of transmit beamforming and phase shifts, which have been passed over the existing works. 

\item To the best of our knowledge, we devise the first \ac{DRL}-based approach for the phase-dependent reflection amplitude model in \ac{RIS}-aided MU-MISO systems to provide an alternative ML-based framework for the suboptimal iterative algorithms proposed in \cite{pdaModel}. 

\item Although the agent is unaware of the phase-dependent reflections and the presence of channel estimation error, the presented method achieves sum rates close to the existing \ac{DRL} agent operating with perfect \ac{CSI} and full awareness of phase-dependent \ac{RIS} amplitudes.

\item To ensure reproducibility and support further research on \ac{DRL}-based \ac{RIS} systems, we provide our source code and results in the GitHub repository\footnote{\url{https://github.com/baturaysaglam/RIS-MISO-PDA-Deep-Reinforcement-Learning}\label{our_repo}}.
\end{itemize}

\section{System Model}
\label{sec:system_model}
We consider the downlink of a narrow-band \ac{RIS}-aided \ac{MU-MISO} system consisting of $K$ single-antenna users, $M$ base station (BS) antennas, and $L$ \ac{RIS} elements. The transmit beamforming matrix $\mathbf{G}\in\mathbb{C}^{M\times K}$ maps $K$ data streams denoted by $\mathbf{x}\in\mathbb{C}^{K\times 1}$ for $K$ users onto $M$ transmit antennas. $\mathbf{H}\in\mathbb{C}^{L\times M}$, $\boldsymbol{\Phi}\triangleq\text{diag}(\phi_1,\dots,\phi_L)\in\mathbb{C}^{L\times L}$, and $\mathbf{h}_k\in\mathbb{C}^{L\times 1}$ denote the \ac{BS}-\ac{RIS} channel, the diagonal reflection matrix at the \ac{RIS}, and \ac{RIS}-user $k$ channel respectively. In the following subsections, we explain two different environment models: \emph{true environment} with phase-dependent \ac{RIS} amplitude and perfect \ac{CSI}, and the \emph{mismatch environment} with ideal reflection assumption and imperfect \ac{CSI}.


\subsection{True Environment Model}
The received signal at user $k$ can be expressed as:
\begin{equation}\label{eq:rxModel}
    z_k = \mathbf{h}_k^\top\boldsymbol{\Phi}\mathbf{H}\mathbf{G}\mathbf{x}+w_k,
\end{equation}
where the complex scalars $z_k$ and $w_k$ denote the received signal and the additive receiver noise at the $k$'th user, respectively, and we assume that $w_k\sim\mathcal{CN}(0,\sigma_w^2)$ for all $k$. The \ac{RIS} follows the phase-dependent amplitude model in \cite{pdaModel}, with entries $\phi_l=\beta(\varphi_l)e^{j\varphi_l}$ for $\:\varphi\in[0,2\pi)$, resulting in:
\begin{equation}\label{eq:pdaModel}
    \beta(\varphi_l) = (1-\beta_{\min})\left(\frac{\sin(\varphi_l-\mu)+1}{2}\right)^{\kappa}+\beta_{\min},
\end{equation}
where $\beta_{\min} \in [0, 1]$, $\mu \geq 0$, and $\kappa \geq 0$ are constants that depend on the hardware implementation of the \ac{RIS}. In the golden standard scenario, the \ac{BS} knows the individual cascaded channels to each user, denoted by:
\begin{equation}\label{eq:cascadedChannel}
    \mathbf{D}_k\triangleq \text{diag}(\mathbf{h}_k)\mathbf{H}\in\mathbb{C}^{L\times M},\quad\forall k=1,\dots,K.
\end{equation}
Hence \eqref{eq:rxModel} can be rewritten as:
\begin{equation}\label{eq:rxModelCascaded}
    z_k = \boldsymbol{\phi}^\top\mathbf{D}_k\mathbf{G}\mathbf{x}+w_k,
\end{equation}
where $\boldsymbol{\phi}\in\mathbb{C}^{L\times 1}$ denotes the column vector consisting of the diagonal entries of $\boldsymbol{\Phi}$.
\subsection{Mismatch Environment Model}
In this simplified model, the \ac{RIS} reflections are assumed to be lossless, i.e., $\boldsymbol{\hat{\phi}}\triangleq[e^{j\varphi_1},\dots,e^{j\varphi_L}]^\top$. Moreover, the agent has access to only an imperfect estimate of the cascaded channels, namely:
\begin{equation}
\label{eq:channel_noise}
    \hat{\mathbf{D}}_k\triangleq \mathbf{D}_k+\mathbf{E}_k,\quad\forall k=1\dots,K,
\end{equation}
where $\mathbf{E}_k\in\mathbb{C}^{L\times M}$ denotes the channel estimation error matrix of the cascaded channel of each user, with \ac{i.i.d.} entries $e_{l,m}^{(k)}\sim\mathcal{CN}(0,\sigma_e^2)$.

\subsection{Problem Formulation}
\label{sec:problem_form}
\subsubsection{The Golden Standard Objective} 
Our emphasis is to utilize the \ac{DRL} agent to maximize the sum downlink rate in the system, which is denoted as:
\begin{align}\label{eq:sumrate}
    R_\Sigma \triangleq \sum_{k=1}^{K}\log\left(1+\frac{\|\boldsymbol{\phi}^\top\mathbf{D}_k\mathbf{G}\|^2}{\sum_{j\neq k}\|\boldsymbol{\phi}^\top\mathbf{D}_j\mathbf{G}\|^2+\sigma_w^2}\right).
\end{align}
The \ac{BS} aims to maximize \eqref{eq:sumrate} by adjusting $\mathbf{G}$ and $\boldsymbol{\phi}$. Under the transmission power constraint $P_t$ and the domain restriction of phase shifts, the optimization problem is expressed as:
\begin{align}\label{eq:optGold}\nonumber
\underset{\boldsymbol{\phi},\mathbf{G}}{\text{maximize}} \quad & R_\Sigma\\\nonumber
\text{subject to} \quad & \varphi_l\in[0,2\pi), \quad \forall \:l=1,\dots,L,\\
&\text{tr}(\mathbf{G}\mathbf{G}^H)\leq P_t.
\end{align}
where $\boldsymbol{\phi}$ depends on $\varphi_1,\dots,\varphi_L$ and $\beta(\varphi_{l})$ according to \eqref{eq:pdaModel} when the \ac{BS} agent is aware of the true environment model.
\subsubsection{The Mismatch Objective} The optimization problem to be solved in the mismatch scenario is defined as:
\begin{align}\label{eq:sumrateMismatch}
    \hat{R}_\Sigma\triangleq \sum_{k=1}^{K}\log\left(1+\frac{\|\boldsymbol{\hat{\phi}}^\top\hat{\mathbf{D}}_k\mathbf{G}\|^2}{\sum_{j\neq k}\|\boldsymbol{\hat{\phi}}^\top\hat{\mathbf{D}}_j\mathbf{G}\|^2+\sigma_w^2}\right).
\end{align}
Consequently, the \ac{BS} agent considers the following optimization problem:
\begin{align}\label{eq:optMismatch}\nonumber
\underset{\boldsymbol{\hat{\phi}},\mathbf{G}}{\text{maximize}} \quad & \hat{R}_\Sigma\\\nonumber
\text{subject to} \quad & \varphi_l\in[0,2\pi), \quad \forall\:l=1,\dots,L,\\
&\text{tr}(\mathbf{G}\mathbf{G}^H)\leq P_t.
\end{align}
The objective in \eqref{eq:sumrateMismatch} uses ideal \ac{RIS} amplitudes and noisy channel estimates instead of the phase-dependent amplitude and the true channel in \eqref{eq:sumrate}. Hence, they have different forms in terms of the transmit beamformer and the \ac{RIS} phase shifts. Consequently, the seemingly similar \eqref{eq:optGold} and \eqref{eq:optMismatch} have different solutions. In other words, the \ac{BS} is trying to solve a different optimization problem than the actual sum rate in the environment, resulting in inferior transmit beamforming and \ac{RIS} configuration designs. In Section \ref{sec:DRL}, we propose a \ac{DRL} framework that overcomes this phenomenon.
Section \ref{sec:DRL}.
\section{The Deep Reinforcement Learning Framework}\label{sec:DRL}
\subsection{Overview}
\label{sec:drl_overview}
At each discrete time step $t$, the agent observes a state $s \in \mathcal{S}$ and takes an action $a \in \mathcal{A}$, and observes a next state $s^{\prime} \in \mathcal{S}$ and receives a reward $r$, where $\mathcal{S}$ and $\mathcal{A}$ are the state and action spaces, respectively. In fully observable environments, the reinforcement learning (RL) problem is usually represented by a Markov decision process, a tuple $(\mathcal{S}, \mathcal{A}, P, \gamma)$, where $P$ is the transition dynamics such that $s^{\prime}, r \sim P(s, a)$ and $\gamma \in [0, 1]$ is a constant discount factor. 

The objective in RL is to find an optimal policy $\pi$ that maximizes the \textit{value} defined by $V_{t} = \sum_{i = 0}^{\infty}\gamma^{i}r_{t + i + 1}$, where the discount factor $\gamma$ prioritizes the short-term rewards. The policy of an agent is regarded as stochastic if it maps states to action probabilities $\pi: \mathcal{S} \rightarrow p(\mathcal{A})$, or deterministic if it maps states to unique actions $\pi: \mathcal{S} \rightarrow \mathcal{A}$. The performance of a policy is assessed under the action-value function (Q-function or critic) that represents $V_{t}$ while following the policy $\pi$ after acting $a$ in state $s$: $Q^{\pi}(s, a) = \mathbb{E}_{\pi}[\sum_{t = 0}^{\infty}\gamma^{t}r_{t + 1}|s_{0} = s, a_{0} = a]$. The Q-function is learned through the Bellman equation \cite{bellman}: 
\begin{equation}
    Q^{\pi}(s, a) = \mathbb{E}_{r, s^{\prime} \sim P, a^{\prime} \sim \pi}[r + \gamma Q^{\pi}(s^{\prime}, a^{\prime})],
\end{equation}
where $a^{\prime}$ is the action selected by the policy on the observed next state $s^{\prime}$.

In deep RL, the critic is approximated by a deep neural network $Q_{\theta}$ with parameters $\theta$, i.e., the Deep Q-learning algorithm \cite{dqn}. Given a transition tuple $(s, a, r, s^{\prime})$, the Q-network is trained by minimizing a loss $J(\theta)$ on the temporal-difference (TD) error $\delta$ corresponding to $Q_{\theta}$ \cite{sutton88}, the difference between the output of $Q_{\theta}$ and learning target $y$:
\begin{gather}
    y \triangleq r + \gamma Q_{\theta^{\prime}}(s^{\prime}, a^{\prime}),\label{eq:q_objective} \\
    \delta \triangleq y - Q_{\theta}(s, a);\label{eq:td_error} \\
    \theta \leftarrow \theta - \eta\nabla_{\theta}J(\theta)\label{eq:q_update},
\end{gather}
where $J(\theta) = \lvert\delta\rvert^{2}$, $\nabla_{\theta}J(\theta)$ is the gradient of the loss $J(\theta)$ with respect to $\theta$, and $\eta$ is the learning rate. The target $y$ in \eqref{eq:q_objective} utilizes a separate target network with parameters $\theta^{\prime}$ that maintains stability and fixed objective in learning the optimal Q-function \cite{dqn}. The target parameters are updated to copy the parameters $\theta$ after a number of learning steps.

\subsection{The Soft Actor-Critic Algorithm}
We use the state-of-the-art \ac{SAC} algorithm \cite{sac} in our work, outperforming prior suboptimal \ac{DRL} algorithms \cite{97,91,93,98} in most \ac{DRL} benchmarks. SAC is an actor-critic, off-policy algorithm that operates in continuous action spaces. It uses a separate actor network to choose actions and stores experiences in the experience replay memory \cite{exp_replay}. Unlike on-policy algorithms, SAC samples transitions from the replay memory for training. Our initial simulations showed that SAC was the only actor-critic algorithm that could converge for the problem of interest despite intensive hyperparameter tuning.

A SAC agent maintains three networks: two Q-networks and a single stochastic policy network (or actor network), each being a multi-layer perceptron (MLP). Using two Q-networks is to reduce the overestimation of Q-value estimates \cite{td3}. The Q-networks take the states provided by the environment and actions produced by the actor network as inputs and produce Q-value estimates, which are scalar values. Given the actor network $\pi_{\psi}$ parameterized by $\psi$, the Q-networks are jointly trained in the SAC algorithm as:
\begin{gather}
    \mathbf{\hat{y}} \triangleq \mathbf{r} + \gamma \underset{i = 1, 2}{\mathrm{min}}Q_{\theta^{\prime}_{i}}(\mathbf{s}^{\prime}, \mathbf{a}^{\prime})\vert_{\mathbf{a}^{\prime} \sim \pi_{\psi}(\cdot | \mathbf{s}^{\prime})} - \alpha\log(\mathbf{a}^{\prime}|\mathbf{s}^{\prime}), \label{eq:modified_q_obj} \\
    J(\theta_{i}) = \frac{1}{N}\|\mathbf{\hat{y}} - Q_{\theta_{i}}(\mathbf{s}, \mathbf{a})\|^{2}_{2},\\
    \theta_{i} \leftarrow \theta_{i} - \eta \nabla_{\theta_{i}}J(\theta_{i})\label{eq:q_network_gradient_descent},
\end{gather}
where $(\mathbf{s}, \mathbf{a}, \mathbf{r}, \mathbf{s}^{\prime})_{i = 1}^{N}$ is the mini-batch of transitions sampled from the experience replay buffer and $N$ is the mini-batch size, $\theta_{i}$ are the parameters corresponding to the $i^{\text{th}}$ Q-network, $\alpha$ is the entropy regularization term, and $\| \cdot \|_{2}$ is the $\text{L}_{2}$ norm. Note that we denote state and action vectors by $s$ and $a$, respectively, while $\mathbf{s}$ and $\mathbf{a}$ represent the batch of state and action vectors.

Similarly, the policy network takes state vectors from the environment as inputs and produces numerical action vectors. The loss for the policy network in the SAC algorithm is expressed by:
\begin{equation}
\label{eq:policy_loss}
    J(\psi) = \frac{1}{N}\sum_{i}^{N}\alpha\log\pi_{\psi}(\mathbf{\hat{a}}_{i}|\mathbf{s}_{i}) - \underset{j = 1, 2}{\mathrm{min}}Q_{\theta_{j}}(\mathbf{s}_{i}, \hat{\mathbf{a}}_{i})\vert_{\hat{\mathbf{a}} \sim \pi_{\psi}(\cdot | \mathbf{s})}.
\end{equation}
Then, the policy gradient $\nabla_{\psi}J(\psi)$ is computed by the stochastic policy gradient algorithm \cite{sac} and used to update the parameters through gradient ascent:
\begin{equation}
\label{eq:policy_gradient_ascent}
    \psi \leftarrow \psi + \eta \nabla_{\psi}J(\psi).
\end{equation}
Lastly, the entropy regularization term $\alpha$ controls exploration, with higher values corresponding to more exploration. While a \ac{DRL} algorithm with a deterministic policy can be used, it requires additive noise for exploration. In contrast, entropy regularization in \ac{SAC} considers the current policy's knowledge, making it a more effective solution for the challenging transmit beamforming and phase shift design.

\subsection{Construction of the Environment}
\label{sec:environment}
RL distinguishes environments into \emph{episodic} and \emph{non-episodic} tasks. An episode ends when a terminal condition is met. In contrast, non-episodic (continuing) tasks have no specific endpoint. In our case, the task is considered continuing because the \ac{BS} continuously performs beamforming and configures \ac{RIS} elements. Terminal conditions can be set, as in \cite{91}, but they might introduce bias and mislead the learning agents \cite{sutton_book}. Therefore, we adopt a continuing task framework in developing our approach.

\subsubsection{Action}
The policy network outputs the flattened concatenation of $\mathbf{G}$ and $\boldsymbol{\phi}$ as the action vector. However, neural networks cannot process complex numbers. Therefore, the actor network produces the real and imaginary parts separately and constructs $\mathbf{G}$ and $\boldsymbol{\phi}$. To satisfy the transmit power and phase domain constraints, i.e., \eqref{eq:optGold} and \eqref{eq:optMismatch}, the agent normalizes the output of actions. Consequently, an action vector consists of $2MK + 2L$ elements. 

\subsubsection{State}
The state vector consists of transmission and reception powers for each user, the previous action, and the cascaded channel matrices $\mathbf{D}_k$ or their estimates $(\hat{\mathbf{D}}_k)$ for $k = 1,\dots,K$, depending on whether the \ac{BS} has perfect or imperfect \ac{CSI}. Similarly, these matrices are flattened and the number of elements is doubled due to the real and imaginary parts except for powers. We consider the transmission powers allocated to each data stream at the BS and the reception power at each user. Consequently, we obtain $2K$ power-related entries. $2KLM$ entries come from the cascaded channel estimates for each user, and $2MK+2L$ entries come from the previous action vector, resulting in a $2KLM+2MK+2L+2K$-dimensional state vector. Furthermore, the correlation between state dimensions degrades the performance of learning RL agents \cite{sutton_book}. Hence, we whiten state vectors after each environment step. Finally, the initial state of the training still requires the action in the previous step. Thus, we initialize $\mathbf{G}$ as an identity matrix and $\boldsymbol{\phi}$ as a vector of ones to constitute the initial environment state. 
\subsubsection{Reward}
At every time step, the reward is determined by the sum downlink rate expressed by either \eqref{eq:sumrate} or \eqref{eq:sumrateMismatch}, depending on the considered objective function.
\section{Methodology}
\subsection{Adapting the Deep Reinforcement Learning Framework to Non-Episodic Tasks}
While $\gamma\ < 1$ prioritizes short-term rewards, the agent must be equally concerned with instantaneous and future rewards since the reward (sum rate) should always be kept maximum \cite{sutton_book}. Therefore, we set $\gamma = 1$. Moreover, agents must carefully remember the outcomes of past actions to compute future actions, which cannot be achieved by learning only from instant rewards \cite{sutton_book}. To solve this, the \ac{DRL} framework should be adapted to non-episodic tasks by considering the \textit{average reward} concept. Therefore, the reward in the current step used to train the agent is modified as follows:
\begin{equation}
\label{eq:avg_reward}
    \Tilde{r} \triangleq r - \Bar{r},
\end{equation}
where $r$ is the instantaneous reward computed by the environment in the current state and $\Bar{r}$ is the average of the rewards collected till the current state. Recall that the definition of $r + Q_{\theta^{\prime}}(s, a)$ (the estimate of the value $Q_{\theta}(s, a)$ when $\gamma = 1$) corresponds to the rewards that the agent will collect till the terminal state. However, there is no terminal state in continuing tasks. Hence, the sum of collected rewards could go to infinity. The average reward overcomes this by restraining the estimation value. Hence, the agent should learn only from $\Tilde{r}$ instead of $r$ or $\Bar{r}$.

\subsection{Maximizing the True Sum Rate Under the Mismatch Environment Model}
To maximize \eqref{eq:sumrate} while learning from \eqref{eq:sumrateMismatch}, we leverage a recent work proposed for the exploration of continuous action spaces, the \ac{DISCOVER} algorithm \cite{batur}. Motivated by the animal psychological systems, \ac{DISCOVER} utilizes a separate \textit{explorer network} $\xi_{\omega}$ with parameters $\omega$ that represents a deterministic \textit{exploration policy}. Its objective is to perturb the actions selected by the policy so that the prediction error by the Q-network is constantly maximized. Consequently, this leads agents to state-action spaces where Q-value prediction is difficult, allowing them to correct the prediction error of unknown or less selected actions. 

However, we do not directly use the \ac{DISCOVER} algorithm since \ac{SAC} already explores the action space by utilizing a stochastic policy with the entropy parameter, i.e., the term $\alpha$ in \eqref{eq:modified_q_obj} and \eqref{eq:policy_loss}. Instead, we slightly modify the DISCOVER algorithm such that the explorer network predicts $\beta(\varphi_{l})$ for $l = 1, \dots, L$ on the observed states:
\begin{equation}
    \xi_{\omega}(s) = 
    \begin{bmatrix}
    \smash{\underbrace{\begin{matrix}
        1& \dots & 1
    \end{matrix}}_{2MK}} & \smash{\underbrace{\begin{matrix}
        \hat{\beta}_{1} & \hat{\beta}_{1} & \hat{\beta}_{2} & \dots & \hat{\beta}_{L} & \hat{\beta}_{L}
    \end{matrix}}_{2L}}
    \end{bmatrix}^{\top},\vspace{9pt}
\end{equation}
where $\hat{\beta}_{l} \in [\beta_{\min}, 1]$. The number of ones in the latter equation is the number of elements included by $\mathbf{G}$ to the action vector. This is feasible since the transmit beamforming produced by the agent does not affect the RIS reflection loss in the phase-dependent amplitude model. In addition, there are two entries for each $\hat{\beta}_{l}$ to scale both the real and imaginary parts of $\boldsymbol{\hat{\phi}}$. Thus, the prediction of $\xi_{\omega}(s)$ is used to perturb only the phase part of the actions:
\begin{equation}
    a_{\hat{\beta}} \triangleq a \odot \lambda \cdot \xi_{\omega}(s) \implies \boldsymbol{\hat{\phi}}_{\hat{\beta}} \triangleq \begin{bmatrix}
    \hat{\beta}_{1}e^{j\varphi_{1}} \dots \hat{\beta}_{L}e^{j\varphi_{L}}
    \end{bmatrix}^\top,
\end{equation}
where $\odot$ is the Hadamard product, and the hyperparameter $\lambda \in (0, 1]$ restricts the explorer network not to perturb the actions chosen by the actor detrimentally, similar to \ac{DISCOVER}. The environment takes $a_{\hat{\beta}}$ from the agent and computes the next state and reward with respect to $a_{\hat{\beta}}$. Therefore, the perturbed actions are sampled from the experience replay buffer instead of the raw ones produced by the policy network. Accordingly, the losses for the Q- and actor networks are modified as follows:
\begin{gather}
   \mathbf{\hat{y}}_{\hat{\beta}} \triangleq \mathbf{\Tilde{r}} + \underset{i = 1, 2}{\mathrm{min}}Q_{\theta^{\prime}_{i}}(\mathbf{s}^{\prime}, \mathbf{a}^{\prime} \odot \lambda \cdot \xi_{\omega^{\prime}}(\mathbf{s})) - \alpha\log(\mathbf{a}^{\prime}|\mathbf{s}^{\prime});\label{eq:modified_y_beta}\\
   J_{\hat{\beta}}(\theta_{i}) \triangleq \frac{1}{N}\|\mathbf{\hat{y}}_{\hat{\beta}} - Q_{\theta_{i}}(\mathbf{s}, \mathbf{a}_{\hat{\beta}})\|^{2}_{2},\label{eq:modified_q_obj_beta} \\ 
   \text{\small$
   J_{\hat{\beta}}(\psi) \triangleq \frac{1}{N}\sum_{i}^{N}\alpha\log\pi_{\psi}(\mathbf{\hat{a}}_{i}|\mathbf{s}_{i}) - \underset{j = 1, 2}{\mathrm{min}}Q_{\theta_{j}}(\mathbf{s}_{i}, \hat{\mathbf{a}}_{\hat{\beta}, i})\vert_{\hat{\mathbf{a}} \sim \pi_{\psi}(\cdot | \mathbf{s})}$\normalsize}\label{eq:modified_policy_loss},
\end{gather}
where $\hat{\mathbf{a}}_{\hat{\beta}, i} = \hat{\mathbf{a}} \odot \lambda \cdot \xi_{\omega}$. Notice that a target explorer network also perturbs the next action $a^{\prime}$, as in DISCOVER. Also, we use $\gamma = 1$ and average reward in \eqref{eq:modified_y_beta}. The explorer network is optimized such that the sum of absolute TD errors is maximized:
\begin{gather}
    \Tilde{\delta}_{\hat{\beta_{i}}}(\mathbf{s}, \mathbf{a} \odot \lambda \cdot \xi_{\omega}(\mathbf{s})) \triangleq \frac{1}{N} \| \mathbf{\hat{y}}_{\hat{\beta}} - Q_{\theta_{i}}(\mathbf{s}, \mathbf{a} \odot \lambda \cdot \xi_{\omega}(\mathbf{s}))\|_{2}^{2},\label{eq:explorer_modified_td_error} \\
     J(\omega) = \Tilde{\delta}_{\hat{\beta_{1}}}(\mathbf{s}, \mathbf{a} \odot \lambda \cdot \xi_{\omega}(\mathbf{s})) + \Tilde{\delta}_{\hat{\beta_{2}}}(\mathbf{s}, \mathbf{a} \odot \lambda \cdot \xi_{\omega}(\mathbf{s}))\label{eq:explorer_loss}.
\end{gather}
The deterministic exploration network is then updated through the Deterministic Policy Gradient algorithm \cite{dpg}:
\begin{gather}
\nabla_{\omega} J(\omega) = \sum_{i = 1}^{2}\mathbb{E}[\nabla_{\zeta}\Tilde{\delta}_{\hat{\beta_{i}}}(\mathbf{s}, \mathbf{a} \odot \zeta)|_{\zeta = \lambda \cdot \xi_{\omega}(\mathbf{s})}\nabla_{\omega}\xi_{\omega}(\mathbf{s})],\label{eq:explorer_policy_gradient} \\
\omega \leftarrow \omega + \eta \nabla_{\omega}J(\omega).\label{eq:explorer_gradient_ascent}
\end{gather}

This forms our framework to solve the downlink RIS-aided MU-MISO system under the phase-dependent amplitude model. Overall, the explorer network predicts $\beta(\varphi_{l})$ and scales the actions selected by the policy using $\hat{\beta}_{l}$. Then, the scaled action is fed to the environment. Notice that the reward (sum rate) computed by the environment is altered with respect to the scaled actions $a_{\hat{\beta}}$ (or $\boldsymbol{\hat{\phi}}_{\hat{\beta}}$):
\begin{equation}
    \hat{R}_{\Sigma, \hat{\beta}} \triangleq \sum_{k=1}^{K}\log\left(1+\frac{\| \boldsymbol{\hat{\phi}}_{\hat{\beta}}^\top\hat{\mathbf{D}}_k\mathbf{G}\|^2}{\sum_{j\neq k}\|\boldsymbol{\hat{\phi}}_{\hat{\beta}}^\top\hat{\mathbf{D}}_j\mathbf{G}\|^2+\sigma_w^2}\right).
\end{equation}
Hence, the agent observes the effect of the explorer network's $\beta(\varphi_{l})$ prediction through the reward it receives, which is equivalent to implicitly learning the true environment model. By maximizing the TD error, the exploration policy further forces the Q-networks to learn from its prediction mistakes since now $\hat{\beta}$-altered actions $a_{\hat{\beta}}$ and rewards $\hat{R}_{\Sigma, \hat{\beta}}$ are included in the loss of Q-networks, i.e., \eqref{eq:modified_q_obj_beta}. Ultimately, the exploration policy learns the true environment model by considering the current knowledge of the Q-networks and policy \cite{batur}. We refer to the resulting algorithm as \textit{$\beta\text{-Space Exploration}$} and provide the pseudocode in our repository\footref{our_repo}. 

\paragraph*{Complexity Analysis} DISCOVER adds another neural network to the training process, slightly increasing computational complexity by less than 33\% compared to SAC's three networks (policy and two critics). The introduced complexity can never be 33\% since the input dimensions of the explorer and actor networks, i.e., only the state dimension, is always less than the input dimension of the critic network, i.e., the aggregated state and action dimensions.

\section{Results} 
\subsection{Simulation Setup}
To test the effectiveness of $\beta\text{-Space Exploration}$, we compare it against two vanilla SAC agents corresponding to two scenarios: golden standard and mismatch. In the golden standard scenario, the agent knows the true environment model and is trained using the rewards computed according to \eqref{eq:sumrate}. Moreover, the agent has perfect \ac{CSI}. In the mismatch case, however, the \ac{BS} tries to solve \eqref{eq:optMismatch}, learns from the rewards computed according to \eqref{eq:sumrateMismatch}, and has imperfect \ac{CSI}. While the vanilla SAC agent is tested under both scenarios, the SAC agent combined with $\beta\text{-Space Exploration}$ is tested under the mismatch scenario.

\begin{table}[!hbp]
\caption{The hyperparameter setting used to produce the reported results. No tuning was performed on the environment parameters.}
\label{table:hyper_param}
\begin{center}
\begin{threeparttable}
    \begin{tabularx}{\linewidth}{l c}
        \toprule
        \textbf{Hyperparameter} & \textbf{Value} \\
        \midrule
        \# of hidden layers\tnote{$\dagger$} & 2 \\
        \# of units in each hidden layer\tnote{$\dagger$} & 256 \\
        Hidden layers activation\tnote{$\dagger$} & ReLU \\
        Final layer activation (Q-networks) & Linear \\
        Final layer activation (actor, explorer) & tanh \\
        Learning rate $\eta$\tnote{$\dagger$} & $10^{-3}$ \\ 
        Weight decay\tnote{$\dagger$} & None \\
        Weight initialization\tnote{$\dagger$} & Xavier uniform \cite{xavier} \\
        Bias initialization\tnote{$\dagger$} & constant \\
        Optimizer\tnote{$\dagger$} & Adam \cite{adam} \\
        Total time steps per training & 20000 \\
        Experience replay buffer size & 20000 \\
        Experience replay sampling method & uniform \\
        Mini-batch size & 16 \\
        Discount term $\gamma$ & 1 \\
        Learning rate for target networks $\tau$\tnote{$\dagger$} & $10^{-3}$ \\
        Network update interval\tnote{$\dagger$} & after each environment step \\
        Initial $\alpha$ & 0.2 \\
        Entropy target & \texttt{-action dimension} \\
        SAC log standard deviation clipping & $(-20, 2)$ \\ 
        SAC $\epsilon$ & $10^{-6}$ \\
        Initial $\beta\text{-Space Exploration}$ $\lambda$ & 0.3 \\
        \midrule
        $\mu$\tnote{$\ddagger$} & 0 \\
        $\kappa$\tnote{$\ddagger$} & 1.5 \\
        Channel noise variance $\sigma_{e}^{2}$\tnote{$\ddagger$} & $10^{-2}$ \\
        AWGN channel variance $\sigma_{w}^{2}$\tnote{$\ddagger$} & $10^{-2}$ \\
        Channel matrix initialization (Rayleigh)\tnote{$\ddagger$} & $\mathcal{CN}(0, 1)$ \\
        \bottomrule
    \end{tabularx}
    \begin{tablenotes}    
        \item[$\dagger$]{Applies to all neural networks}
        \item[$\ddagger$]{Environment hyperparameter}
    \end{tablenotes}
\end{threeparttable}
\end{center}
\end{table}

\begin{figure*}[!htp]
    \captionsetup[subfloat]{justification=centering}
    \centering
    \begin{equation*}
        \text{{\orange} SAC (golden standard)} \qquad \text{{\blue} SAC (mismatch)} \qquad \text{{\green} SAC + $\beta\text{-Space Exploration}$ (mismatch)}
    \end{equation*}
	\subfloat[$\beta_{\min}$ = 0.3, $P_{t}$ = 30 dBm, \newline $K$ = 4, $M$ = 4, $L$ = 16\label{fig:sc_a}]{
		\includegraphics[width=1.65in, keepaspectratio]{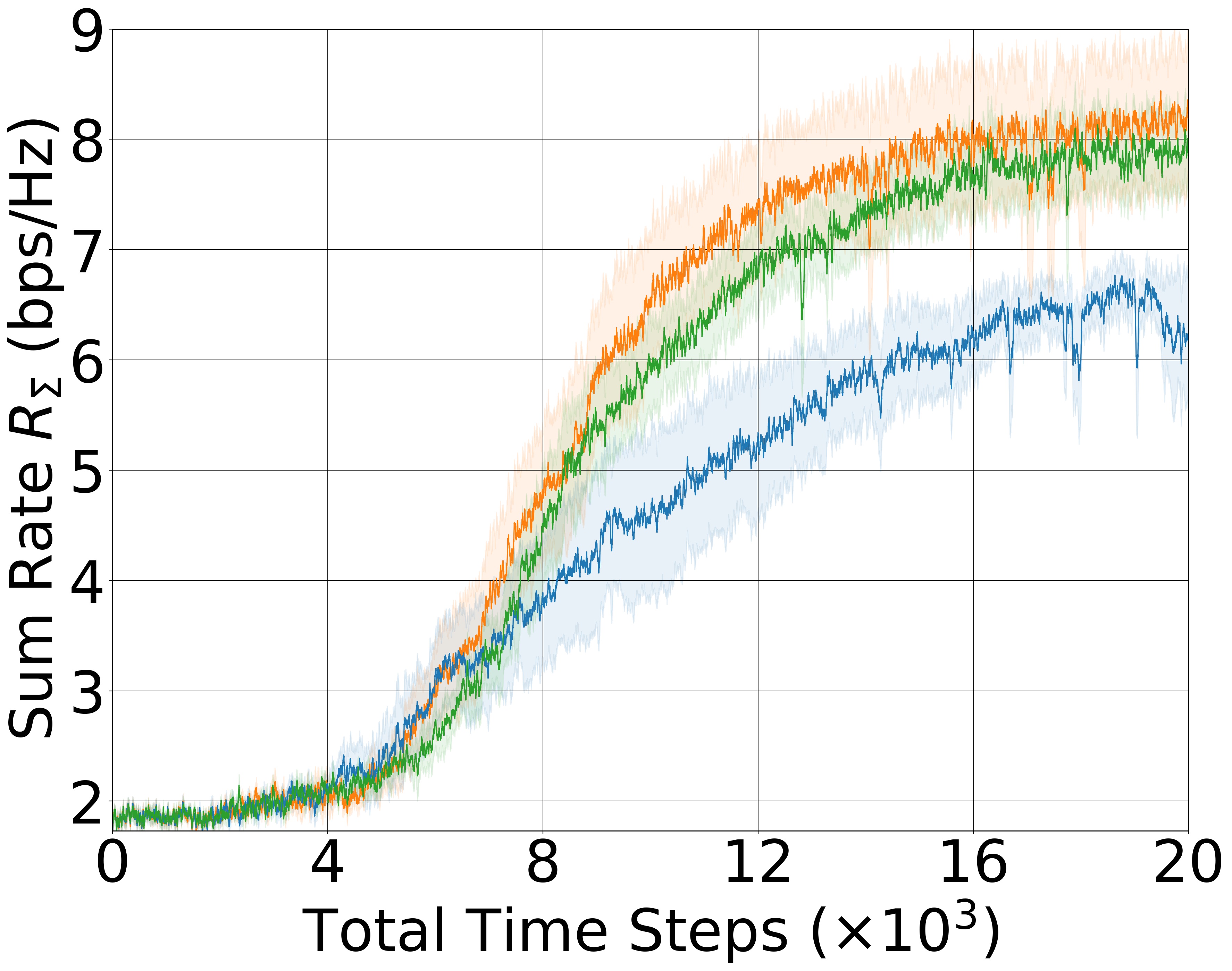}
	}
	\subfloat[$\beta_{\min}$ = 0.6, $P_{t}$ = 30 dBm, \newline $K$ = 4, $M$ = 4, $L$ = 16\label{fig:sc_b}]{
		\includegraphics[width=1.65in, keepaspectratio]{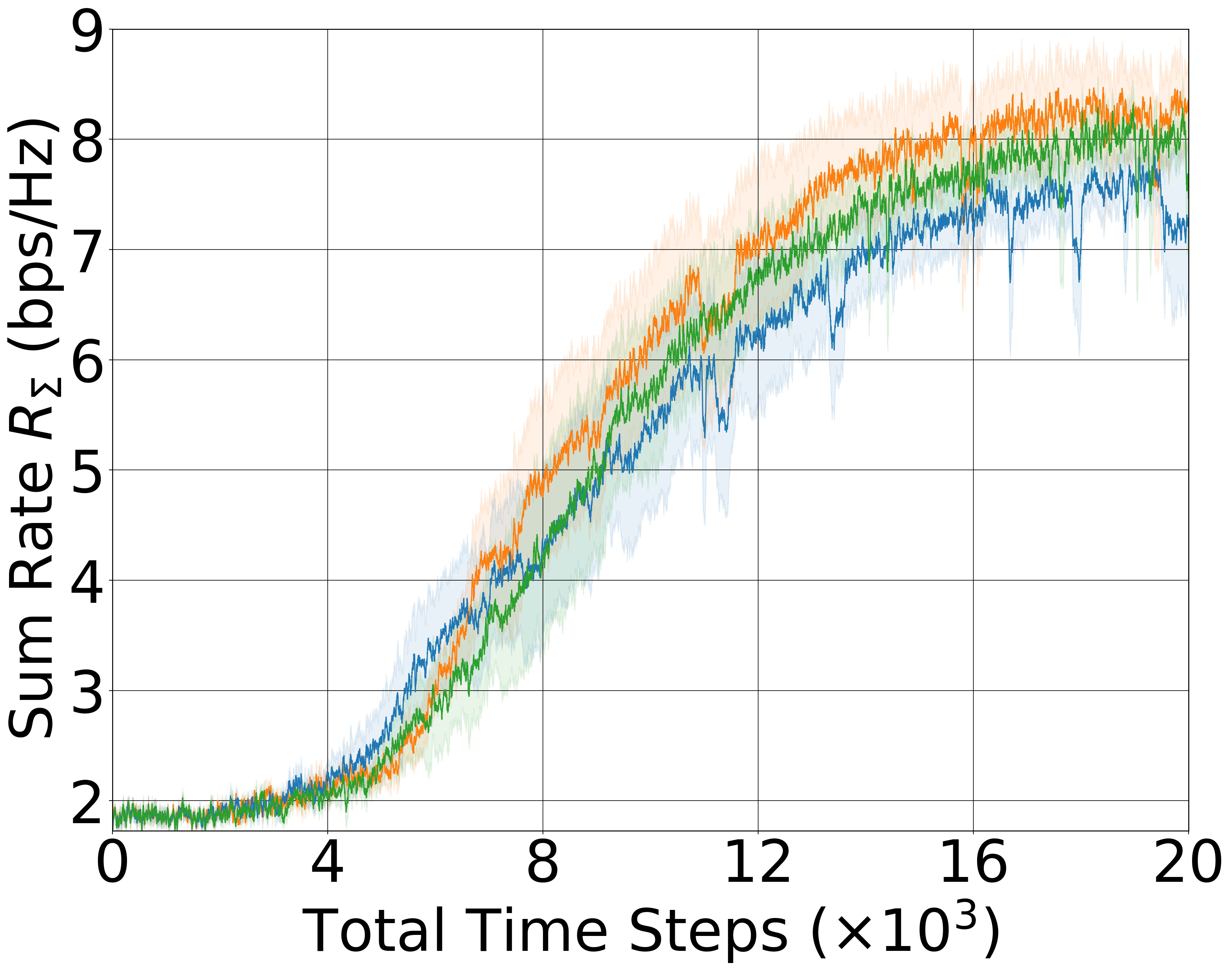}
	}
	\subfloat[$\beta_{\min}$ = 0.6, $P_{t}$ = 30 dBm, \newline $K$ = 4, $M$ = 4, $L$ = 64\label{fig:sc_c}]{
		\includegraphics[width=1.65in, keepaspectratio]{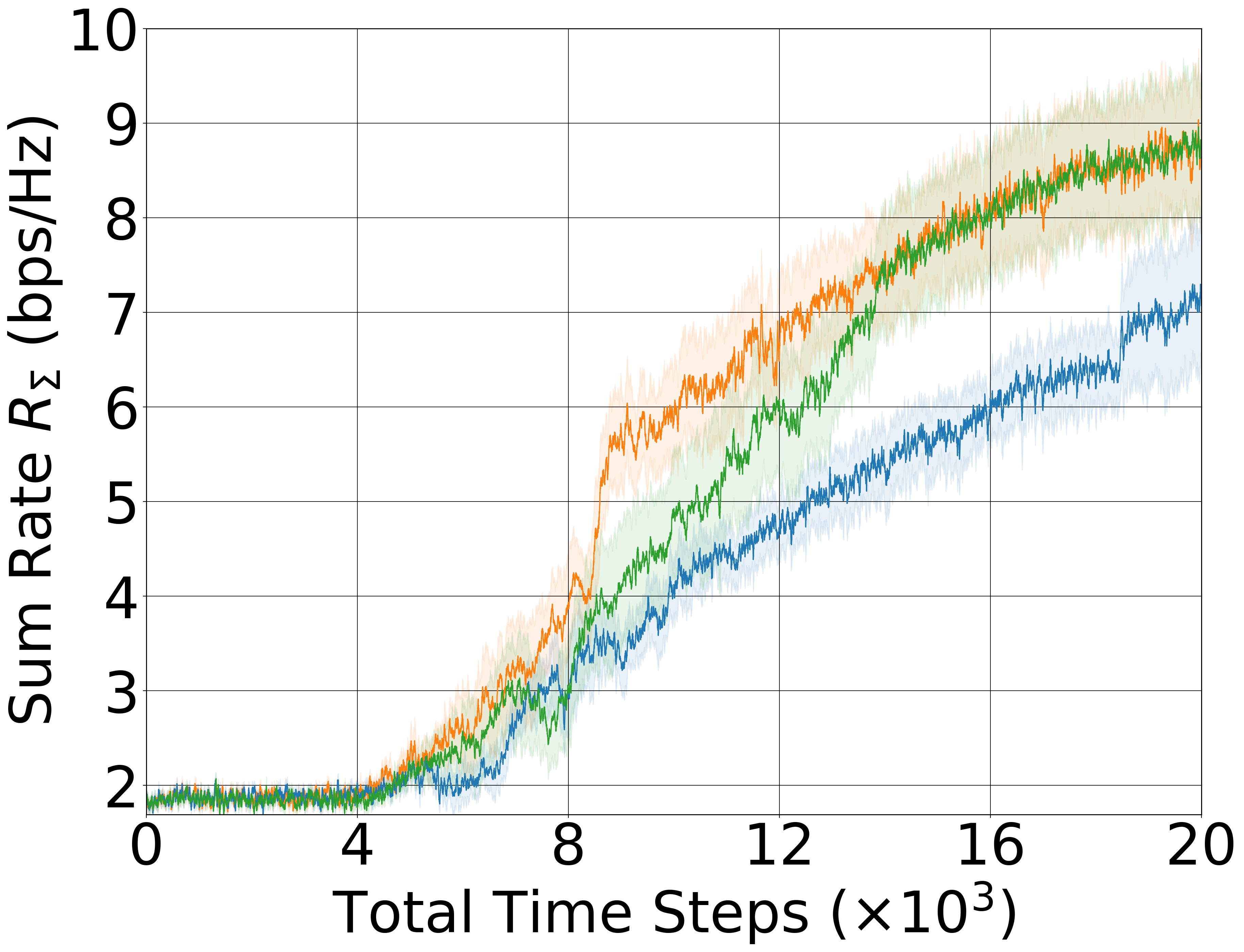}
	}
	\subfloat[$\beta_{\min}$ = 0.6, \newline $P_{t}$ = $\{5, 10, 15, 20, 25, 30\}$ dBm, \newline $K$ = 4, $M$ = 4, $L$ = 16\label{fig:power}]{
		\includegraphics[width=1.65in, keepaspectratio]{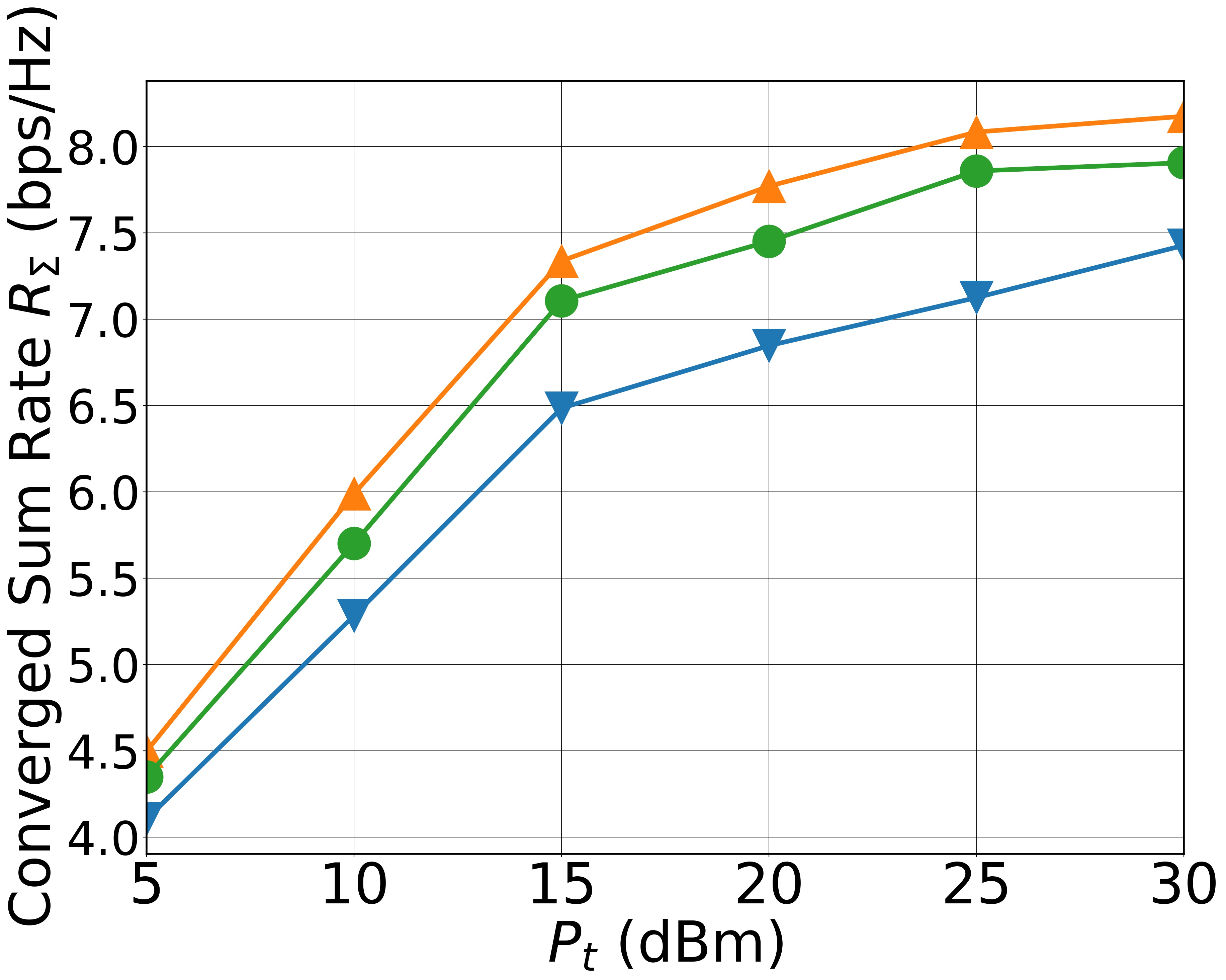}
	} \\
	\caption{Learning curves for the tested settings. Shaded regions represent 95\% confidence intervals over 10 random seeds for each result. A sliding window of size 25 smooths the curves for visual clarity.}
	\label{fig:results}
\end{figure*}

\begin{table*}[!hb]
\begin{center}
\caption{Average of last 1000 instant rewards achieved by the SAC agents, computed according to \eqref{eq:sumrate}, over 10 trials of 20000 time steps. $\pm$ captures a 95\% confidence interval over the trials. The performance increase denotes the percentage of mean sum rate improvement obtained by $\beta\text{-Space Exploration}$ over the vanilla SAC agent in the mismatch environment with respect to the difference between the golden standard and mismatch scenarios.}
\label{table:results}
\begin{tabular}{l c c c c}
    \toprule
    \textbf{Setting} & \textbf{Golden Standard} & \textbf{Mismatch} & \textbf{$\beta\text{-Space Exploration}$} & \textbf{Performance Increase} \\
    \midrule
    $\beta_{\min}$ = 0.3, $P_{t}$ = 30 dBm, $K$ = 4, $M$ = 4, $L$ = 16 & 8.16 $\pm$ 1.24 & 6.36 $\pm$ 0.67 & 7.88 $\pm$ 0.69 & 84\% \\
    $\beta_{\min}$ = 0.6, $P_{t}$ = 30 dBm, $K$ = 4, $M$ = 4, $L$ = 16 & 8.18 $\pm$ 0.77 & 7.43 $\pm$ 0.78 & 7.91 $\pm$ 0.43 & 64\% \\
    $\beta_{\min}$ = 0.6, $P_{t}$ = 30 dBm, $K$ = 4, $M$ = 4, $L$ = 64 & 8.71 $\pm$ 0.84 & 7.00 $\pm$ 0.57 & 8.70 $\pm$ 0.94 & 99\% \\ \\
    
    $P_{t}$ = 5\hphantom{0} dBm, $\beta_{\min}$ = 0.6, $K$ = 4, $M$ = 4, $L$ = 16 & 4.50 $\pm$ 0.45 & 4.11 $\pm$ 0.29 & 4.35 $\pm$ 0.36 & 62\% \\
    $P_{t}$ = 10 dBm, $\beta_{\min}$ = 0.6, $K$ = 4, $M$ = 4, $L$ = 16 & 5.99 $\pm$ 0.41 & 5.28 $\pm$ 0.42 & 5.70 $\pm$ 0.39 & 59\% \\
    $P_{t}$ = 15 dBm, $\beta_{\min}$ = 0.6, $K$ = 4, $M$ = 4, $L$ = 16 & 7.34 $\pm$ 0.83 & 6.49 $\pm$ 0.61 & 7.11 $\pm$ 0.54 & 73\% \\
    $P_{t}$ = 20 dBm, $\beta_{\min}$ = 0.6, $K$ = 4, $M$ = 4, $L$ = 16 & 7.77 $\pm$ 0.60 & 6.85 $\pm$ 0.77 & 7.45 $\pm$ 0.50 & 65\% \\
    $P_{t}$ = 25 dBm, $\beta_{\min}$ = 0.6, $K$ = 4, $M$ = 4, $L$ = 16 & 8.08 $\pm$ 0.83 & 7.13 $\pm$ 0.71 & 7.86 $\pm$ 0.57 & 77\% \\
    $P_{t}$ = 30 dBm, $\beta_{\min}$ = 0.6, $K$ = 4, $M$ = 4, $L$ = 16 & 8.18 $\pm$ 0.77 & 7.43 $\pm$ 0.78 & 7.91 $\pm$ 0.43 & 64\% \\
    \bottomrule
\end{tabular}
\end{center}
\end{table*}

Our simulations follow the well-known \ac{DRL} benchmarking standards \cite{deep_rl_that_matters}, that is, each experiment runs over ten random seeds for a fair comparison with the baselines. Furthermore, the implementation of the SAC algorithm follows the structure outlined in the original paper \cite{sac}. We performed an extensive hyperparameter tuning starting from the hyperparameter setting provided by \cite{sac}. The tuned hyperparameter setting is outlined in Table \ref{table:hyper_param} along with the chosen environment parameter values. We also linearly decay the exploration regularization term $\lambda$ such that it becomes zero at the end of the training. Highly perturbed actions (i.e., large $\lambda$ values) in the final steps may degrade the performance of a SAC agent that learned to control the environment sufficiently well. Precise experimental setup and implementation can be found in the code of our repository\footref{our_repo}.

\subsection{Discussion}
In Fig. \ref{fig:results}, we report the instantaneous sum rates computed according to \eqref{eq:sumrate}, averaged over ten random seeds. While the agents are trained using the average reward $\Tilde{r}$ in \eqref{eq:avg_reward}, the performance is assessed under the instantaneous rewards $r$. Also, Table \ref{table:results} reports the converged sum rates, being the average of the last 1000 instant rewards over ten trials, per the \ac{DRL} benchmarking standards \cite{deep_rl_that_matters}.

From the evaluation results, we infer that $\beta\text{-Space Exploration}$ attains near-optimal results in all of the settings tested. Specifically, when $\beta_{\min} = 0.3$, the SAC agent under the mismatch environment shows considerably worse performance than the golden standard. The resulting performance approaches the golden standard when $\beta_{\min}$ is increased to $0.6$. This is expected since the interval for possible \ac{RIS} loss factors shrinks as $\beta(\varphi_{l}) \in [\beta_{\min}, 1]$. However, our method is not affected by the $\beta_{\min}$ value. For each value of $\beta_{\min}$, it exhibits a robust performance, achieving high sum downlink rates slightly lower than the golden standard. Furthermore, $\beta\text{-Space Exploration}$ regards no issues with the convergence rate, that is, learning curves are practically parallel to the golden standard. This implies that the exploration policy can implicitly learn how its action selections affect the loss in the \ac{RIS} reflections, which is done in a negligible amount of time compared to the total training duration.

When the number of \ac{RIS} elements $L$ is increased to 64, the sum rate achieved by the golden standard  increases slightly due to additional degrees of freedom to control the propagation environment. On the other hand, the vanilla SAC agent cannot benefit from this due to the increased number of misspecified \ac{RIS} amplitudes. In contrast, our proposed method converges to the sum rates achieved by the golden standard with a slight delay, despite being trained according to \eqref{eq:sumrateMismatch}. As shown in Table \ref{table:results}, $99\%$ of the sum rate loss caused by the mismatch is compensated by $\beta$-Space Exploration when $L=64$. We also observe that $\beta\text{-Space Exploration}$ offers consistent performance gains over the vanilla SAC agent under mismatch for different transmit power levels. Additionally, the resulting confidence intervals of our algorithm are usually tighter than the ones corresponding to the golden standard. This suggests that our framework improves credibly over the baseline due to the structure of the introduced method rather than unintended consequences or any exhaustive hyperparameter tuning. 

Lastly, the computational cost of the Q- and actor networks of the SAC algorithm in the considered environment depend on the values of the environment setting parameters $M$, $K$, and $L$. Increasing these parameters increases the number of state and action dimensions, which in turn increases the number of parameters and operations involved in the forward pass of the networks, leading to a higher computational cost. Therefore, the values of $M$, $K$, and $L$ should be chosen carefully to balance the performance of the selected \ac{DRL} algorithm with its computational efficiency.

\section{Concluding Remarks}
In this paper, we present a novel \ac{DRL}-based approach, $\beta$-Space Exploration, to address the three critical aspects of non-episodic tasks, imperfect \ac{CSI}, and hardware impairments in \ac{RIS}-aided \ac{MU-MISO} systems represented by the phase-dependent reflection amplitude model \cite{pdaModel}. Our method jointly designs transmit beamforming and phase shifts to maximize the sum downlink rate of the users. The empirical studies show that $\beta$-Space Exploration attains near-optimal results, is robust to various settings, and compensates for the sum rate loss caused by hardware impairments in the \ac{RIS}. Consequently, our findings highlight the potential of our approach as a promising solution to overcome hardware impairments in \ac{RIS}-aided wireless communication systems. In addition, while the current work considers slow-fading channels, channel aging models can easily be added to our environment code although there exist many opportunities to improve the \ac{DRL} agent design with channel aging in mind.

\bibliographystyle{IEEEtran}
\bibliography{references}

\end{document}